\documentclass[english,aps,twocolumn]{revtex4-1}
\usepackage[T1]{fontenc}
\usepackage[latin9]{inputenc}
\usepackage{amsmath}
\usepackage{graphicx}
\usepackage{appendix}

\makeatletter

\usepackage{babel}

\makeatother

\usepackage{babel}
\begin{document}
\title{A delay differential eqution NOLM-NALM mode-locked laser model}
\author{A. G. Vladimirov$^{a}$, S. Suchkov$^{b}$, G. Huyet$^{c}$, S. K.  Turitsyn$^{b,d}$}
\affiliation{$^{a}$Weierstrass Institute, Mohrenstr. 39, 10117 Berlin, Germany}
\affiliation{$^{b}$Aston-NSU International Centre for Photonics, Novosibirsk State University, Novosibirsk,
Russia}
\affiliation{$^{c}$Universit{\'e} C{\^o}te d\textquoteright Azur, Centre National de La
Recherche Scientifique, Institut de Physique de Nice, F-06560 Valbonne,
France}
\affiliation{$^{d}$Aston Institute of Photonic Technologies, Aston University, B4 7ET, Birmingham, UK}
\begin{abstract}
Delay differential equation model of a NOLM-NALM mode-locked laser is
developed that takes into account finite relaxation rate of the gain
medium and asymmetric beam splitting at the entrance of the nonlinear
mirror loop. Asymptotic linear stability analysis of the continuous wave solutions
performed in the limit of large delay indicates that in a class-B
laser flip instability leading to a period doubling cascade and development of square-wave patterns can be suppressed by a short wavelength modulational instability. Numerically it is shown that the model can demonstrate large windows of regular fundamental and harmonic mode-locked regimes with
single and multiple pulses per cavity round trip time separated by domains of irregular pulsing.
\end{abstract}
\maketitle

\section{Introduction}

Passively mode-locked lasers have attracted much attention in the
recent decades, due to their numerous applications in science, biomedicine, and industry. In passive mode-locking the presence of saturable absorption in the laser cavity is needed to allow short pulse generation. Among different mechanisms to create saturable absorption, a promising one relies on the use of the nonlinear optical/amplifying loop mirror
(NOLM/NALM) \cite{doran1988nonlinear}, which contains a bidirectional loop with asymmetrically located absorber/gain medium and nonlinear element. These configurations are also known as figure-eight lasers \cite{F801,F802}. As a result of the interference of two counterpropagating waves the reflectivity of the NOLM-NALM depends strongly on the power of the incident beam, which creates an effective saturable absorption mechanism.

Most of the models used for theoretical analysis of NOLM-NALM mode-locked lasers are based on the NLS- and Ginzburg-Landau-type equations (see, e.g. \cite{theimer1997figure,salhi2008theoretical,li2016characterization,smirnov2017layout,cai2017state,boscolo2019performance,deng2020energy} and references therein),
where the dynamics of the gain is determined by the average
intra-cavity laser power. An alternative
approach to the modeling of NOLM-NALM lasers was proposed in \cite{vladimirov2019dynamics}
where a simple delay differential equation (DDE) model of a nonlinear mirror mode-locked laser was developed using the approach of Refs. \cite{VTK,VT04,VT05} and analyzed analytically and numerically. Later a similar DDE model was used in \cite{aadhi2019highly} to describe a mode-locking in a NALM
mode-locked laser with a semiconductor optical amplifier (SOA) in
the nonlinear mirror loop. Both these models, however, assume adiabatic elimination of the inversion in the gain medium and symmetric beam splitter connecting the main laser cavity with the nonlinear mirror loop. On the other hand,
asymmetric beam splitters are widely used in nonlinear mirror mode-locked lasers, see e.g. \cite{theimer1997figure,tran2008switchable,yang2012chaotic,yun2012observation,li2014all}.
Furthermore, as soon as the pulse duration becomes smaller that the gain relaxation time the effect of the gain dynamics on the pulse shaping must be taken into account \cite{nizette2021generalized}.
An empirical NOLM model including a rate equation for the population inversion dynamics was reported in \cite{yang2012chaotic}. However, since such important physical factor as spectral filtering of the laser radiation is missing in this model, similarly to the Poincare map model used in \cite{lai2005nolm}, it is hardly applicable to describe
short pulse generation in mode-locking regimes. The aim of this paper is to generalize the model developed in \cite{vladimirov2019dynamics}
to the case of arbitrary population relaxation rates of the laser gain medium and beam splitting ratios. Note that similarly to the models discussed in \cite{vladimirov2019dynamics,aadhi2019highly} our model assumes that chromatic dispersion of the intracavity medium does not play an important role in the mechanism of the mode-locked pulse formation. Although the chromatic dispersion can be included into DDE models \cite{pimenovprl,pimenov2020temporal}, this task is beyond the scope of the present work.

Using the generalized model we investigate analytically the stability and bifurcations of continuous wave (CW) solutions in the limit of large delay. Numerical simulations reveal large domains of fundamental and harmonic mode-locking regimes in the parameter space. We show that both the inversion relaxation rate and beam splitting ratio can strongly affect the dynamics of the system and the existence domains of stable mode-locked regimes.

\section{Model equations}

\begin{figure}
\includegraphics[scale=0.6]{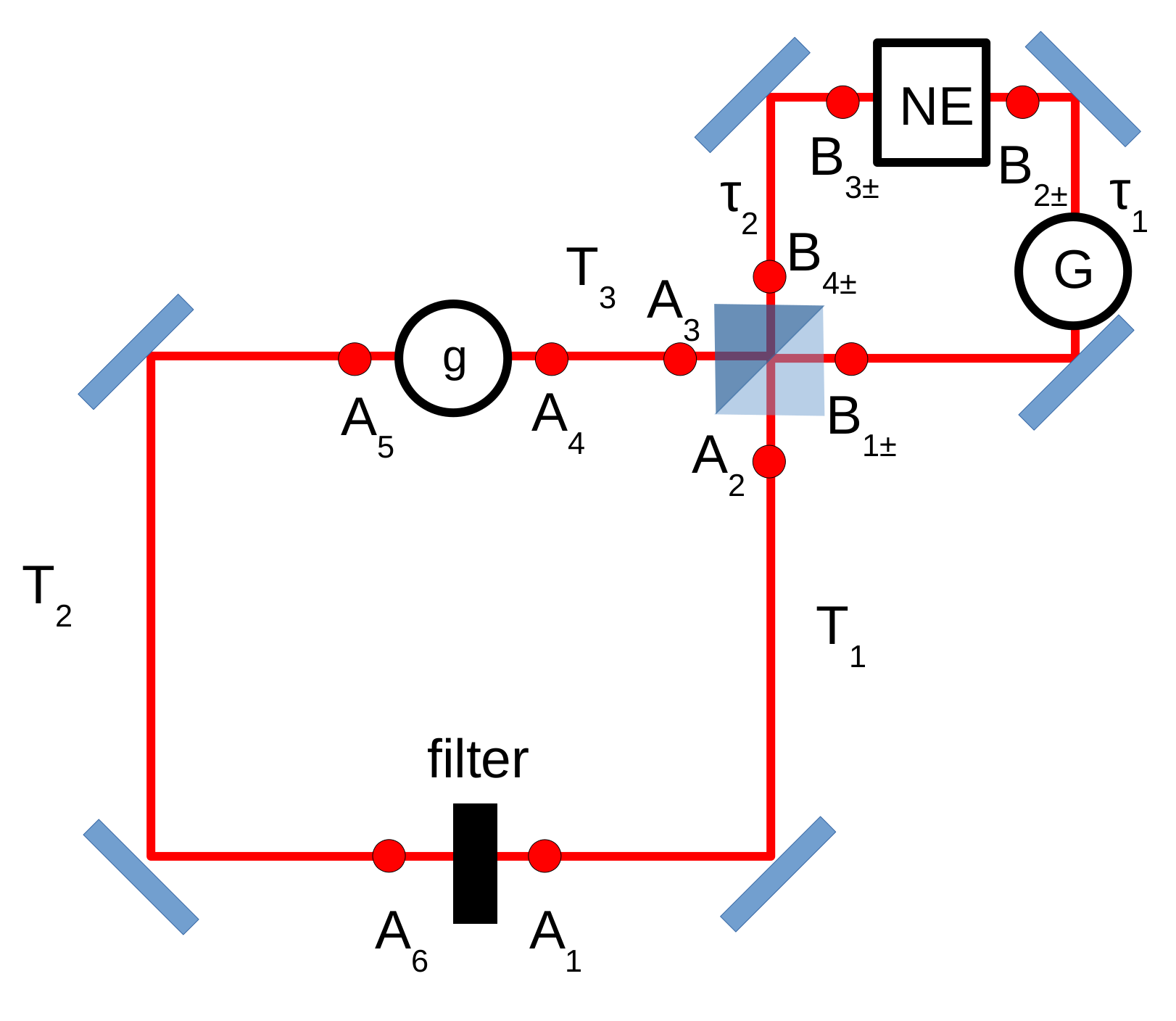}

\caption{Schematic representation of the NOLN-NALM laser. g -- amplifying
medium in the main cavity, G -- linear loss/gain in the nonlinear
mirror loop, NE -- nonlinear element, $A_{1,2,3,4,5,6}$ -- electric
field envelopes in the main cavity (unidirectional, counterclockwise),
$B_{\pm1,\pm2,\pm3,\pm4}$ -- electric field envelopes in the nonlinear
mirror loop (bidirectional). Here the subscripts ``$+$'' and ``$-$''
correspond to counterclockwise and clockwise waves, respectively.\label{fig:Schematic}}
\end{figure}

In this section we extend the NOLM-NALM mode-locked laser model proposed
by the last author of \cite{vladimirov2019dynamics} to the case
of finite carrier relaxation rate and arbitrary beam spliting ratio.
A schematic presentation of the laser system under consideration is
given in Fig. \ref{fig:Schematic}.

To derive our model equations we use the lumped element approach similar to that described in \cite{VTK,VT04,vladimirov}. Propagation of the electric field envelope in the passive sections
of the main unidirectional cavity can be described by the relations:
\begin{multline}
A_{2}\left(t\right)=\sqrt{\kappa_{2}}A_{1}\left(t-T_{1}\right),\quad A_{4}\left(t\right)=A_{3}\left(t-T_{3}\right),\\
A_{6}\left(t\right)=\sqrt{\kappa_{1}}A_{5}\left(t-T_{2}\right),\label{eq:empty}
\end{multline}
where $0<\kappa_{1,2}<1$ are the intensity attenuation factors due to the linear non-resonant losses in the intracavity media and output of radiation through the mirrors. $T_{1,2}$ are the delay times introduced by the passive sections. 

The gain section of
the main cavity can be described by the relation:
\begin{equation}
A_{5}\left(t\right)=A_{4}\left(t-T_{4}\right)e^{\frac{1-i\alpha}{2}g\left(t\right)},\label{eq:SOA}
\end{equation}
where $\alpha$ is the linewidth enhancement factor, which is nonzero in the case when SOA is used as a gain medium. Experimental study of a NOLM mode-locked laser with SOA amplifying medium was reported in
\cite{cai201040}. The time evolution of the cumulative gain $g\left(t\right)$
is governed by ordinary differential equation \cite{agrawal1989self,VT05}
\begin{equation}
\gamma^{-1}\frac{dg}{dt}=p-g-\left(e^{g}-1\right)\left|A_{4}\left(t-T_{4}\right)\right|^{2}.\label{eq:gain}
\end{equation}
Here $\gamma$ is the normalized carrier relaxation rate, and $p$
is the linear gain (pump) parameter. 

The transformations of the counterpropagating field amplitudes by
the linear gain (loss) and passive sections of the nonlinear
mirror loop are given by
\begin{multline}
B_{-1}\left(t\right)=\sqrt{G}B_{-2}\left(t-\tau_{1}\right),\quad B_{+2}=\sqrt{G}B_{+1}\left(t-\tau_{1}\right),\\
B_{-1}\left(t\right)=B_{-2}\left(t-\tau_{2}\right),\quad B_{+4}\left(t\right)=B_{+3}\left(t-\tau_{2}\right),\label{eq:G}
\end{multline}
where $G>1$ ($G<1$) correspond to NALM (NOLM), the subscript ``$+$''  (``$-$'') denotes counterclockwise (clockwise) propagating wave, and $\tau_{1,2}$ are the corresponding delay times, which depend
on the length of the passive sections of the nonlinear mirror loop. For the Kerr element insede this loop we can write 
\begin{eqnarray}
B_{-2}\left(t\right) & = & B_{-3}\left(t-\tau_{3}\right)e^{-ia\left[\left|B_{-3}\left(t-\tau_{3}\right)\right|^{2}+h\left|^{2}B_{+2}\left(t-\tau_{3}\right)\right|\right]},\nonumber \\
B_{+3}\left(t\right) & = & B_{+2}\left(t-\tau_{3}\right)e^{-ia\left[\left|B_{+2}\left(t-\tau_{3}\right)\right|^{2}+h\left|B_{-3}\left(t-\tau_{3}\right)\right|^{2}\right]},\label{eq:NEcubic}
\end{eqnarray}
where $a$ and $\tau_{3}$ are the Kerr coefficient and
time delay introduced by the nonlinear element, respectively. Both
these quantities are proportional to the length of the nonlinear element.
The parameter $h$ is responsible for the standing wave effect. Since in mode-locking regime when the pulse duration is much smaller then the cavity round trip time one can neglect the interference of the two counter-propagating pulses in the nonlinear element, below we assume
that $h=0$ in Eq. (\ref{eq:XYcubic}). 

The beam splitter with $K:1-K$ intensity ratio is described by
\begin{multline}
B_{+1}\left(t\right)=-\sqrt{K}A_{2}\left(t\right),\quad B_{-4}\left(t\right)=\sqrt{1-K}A_{2}\left(t\right),\\
A_{3}\left(t\right)=\sqrt{1-K}B_{-1}\left(t\right)+\sqrt{K}B_{+2}\left(t\right),\label{eq:splitter}
\end{multline}
where $0<K<1$ and the sign ``$-$'' in the first equation corresponds
to the reflection from a more dense medium. $K=0.5$ corresponds to
symmetric 50:50 splitter.

Finally, thin Lorentzian spectral filtering element is described by: 
\begin{equation}
\Gamma^{-1}\frac{dA_{1}\left(t\right)}{dt}+A_{1}\left(t\right)=A_{6}\left(t\right).\label{eq:filter}
\end{equation}
 Below we assume that the time $t$ is normalized in such a way that
the spectral width of the filter is $\Gamma=1$. In this case the inversion relaxation rate in Eq. (\ref{eq:gain}) is normalized by the spectral filter width $\gamma=(\tau_g\Gamma)^{-1}$, where $\tau_g$ is the dimensional inversion relaxation time. Note that the inverse filter spectral width $\Gamma^{-1}$ gives approximately the lower limit for the pulse width $\tau_p$ generated by the NOLM-NALM mode-locked laser. Therefore, we get the relation $\gamma\gtrsim\tau_p/\tau_g$.

\begin{figure}
\includegraphics[scale=0.4]{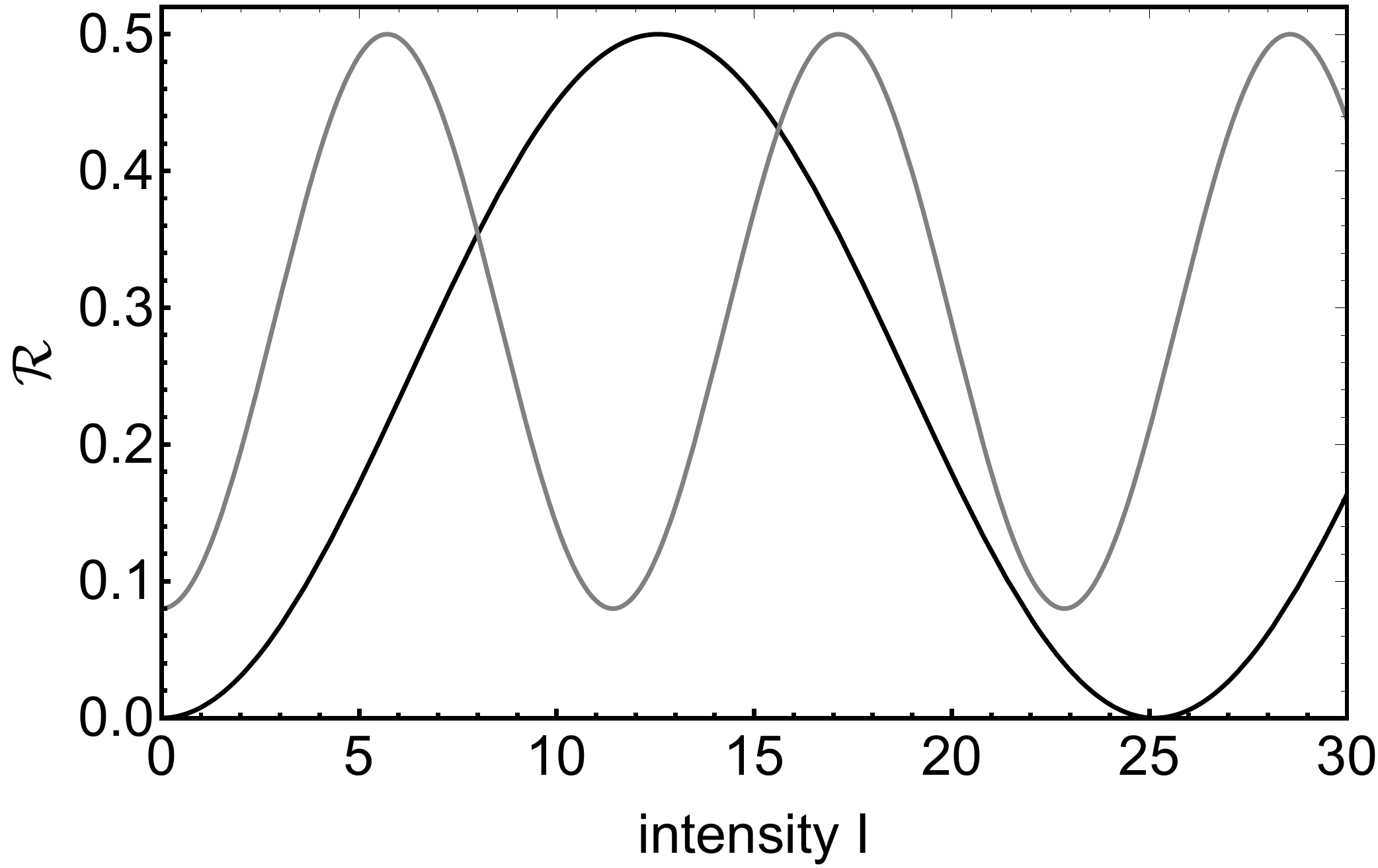}
\caption{\label{fig:reflectivity}Intensity reflectivity ${\cal R}$ of the nonlinear mirror defined by Eq. (\ref{eq:transm}) as a function of
the field intensity $I=\left|A\right|^{2}$. Black (gray) line corresponds to symmetric (asymmetric) beam splitter, $K=0.5$ ($K=0.3)$. 
Other parameters are $a=1.0$ and $G=0.5$.}
\end{figure}

Substituting the relations (\ref{eq:empty})-(\ref{eq:filter}) into
one another we get our master NOLM-NALM laser model:

\begin{equation}
\frac{dA}{dt}+A=\sqrt{\kappa}e^{(1-i\alpha)g/2+i\theta}r\left(\left|A_{T}\right|^{2}\right)A_{T},\label{eq:Model1}
\end{equation}
\begin{equation}
\gamma^{-1}\frac{dg}{dt}=p-g-\left(e^{g}-1\right)|A_{T}|^{2}\left|r\left(\left|A_{T}\right|^{2}\right)\right|^{2},\label{eq:Model2}
\end{equation}
where $A\equiv A_{1}$ , $\kappa=\kappa_{1}\kappa_{2}$ describes
the total linear non-resonant losses in the main cavity per round trip,
$0<\kappa<1$, the subscript $T$ denotes the delayed argument, $A_{T}=A\left(t-T\right)$
with the delay time $T=T_{1}+T_{2}+T_{3}+T_{4}+\tau_{1}+\tau_{2}+\tau_{3}$
equal to the normalized cold cavity round trip time, and the complex nonlinear mirror amplitude reflection coefficient $r$ is given by 
\[
r\left(\left|A_{T}\right|^{2}\right)=\sqrt{G}\left[\left(1-K\right)e^{iX_{T}}-Ke^{iY_{T}}\right]
\]
with 
\begin{equation}
X_{T}=-a\left[\left(1-K\right)\left|A_{T}\right|^{2}\right],\quad Y_{T}=-a\left[KG\left|A_{T}\right|^{2}\right].\label{eq:XYcubic}
\end{equation}
The intensity reflectivity coefficient ${\cal R}$ ($0\le{\cal R}\leq G$)
of the nonlinear mirror is given by 
\begin{multline}
{\cal R}\left(\left|A\right|^{2}\right)=\left|r\left(\left|A\right|^{2}\right)\right|^{2}\\
=G\left(1-2K\left(1-K\right)\left\{ 1+\cos\left(a\left(1-K-GK\right)\left|A\right|^{2}\right)\right\} \right)\\
=G\left(1-4K\left(1-K\right)\cos\left[\frac{a}{2}\left(1-K-GK\right)\left|A\right|^{2}\right]^{2}\right).\label{eq:transm}
\end{multline}
This coefficient coincides with that
reported in Refs. \cite{doran1988nonlinear,lai2005nolm,fermann1990nonlinear}.
The dependence of the coefficient ${\cal R}$ on the laser field intensity
$I=\left|A\right|^{2}$ is illustrated in Fig. \ref{fig:reflectivity}
for the case of 50:50 and 30:70 beam splitter. It is seen that
for symmetric beam splitting the reflectivity is zero at $I=0$ and
oscillates with the intensity between $0$ and $G$ with the period $4\pi/\left[a\left(1-K-GK\right)\right]$. If on the other hand the beam splitter is asymmetric,
minimal reflectivity of the nonlinear mirror becomes greater than,
 ${\cal R}_{min}=G\left(1-2K\right)^{2}>0$.

\section{CW regimes}

Trivial solution of Eqs. (\ref{eq:Model1}) and (\ref{eq:Model2})
corresponding to laser off regime is given by $A=0$ and $g=p$. This
solution is stable below the linear threshold defined by $\kappa{\cal R}\left(0\right)e^{p}=1$,
where ${\cal R}\left(0\right)=G\left(1-2K\right)^{2}$. It is seen
from this expression that the threshold value of the pump parameter
$p=p_{0}$ is minimal for $K=0$ and $K=1$, $p_{0}=-\ln\left(\kappa G\right)$,
and tends to infinity for $K\to0.5$. This means that for symmetric
beam splitter laser off solution is always stable \cite{vladimirov2019dynamics}.
Note that since linear gain in the nonlinear mirror loop cannot exceed the total losses in the cavity the product $\kappa G$ should be less
than unity. All our calculations below are performed for the case
of NOLM when $G<1$ and the condition $\kappa G<1$ is satisfied automatically.

\begin{figure}
\includegraphics[scale=0.45]{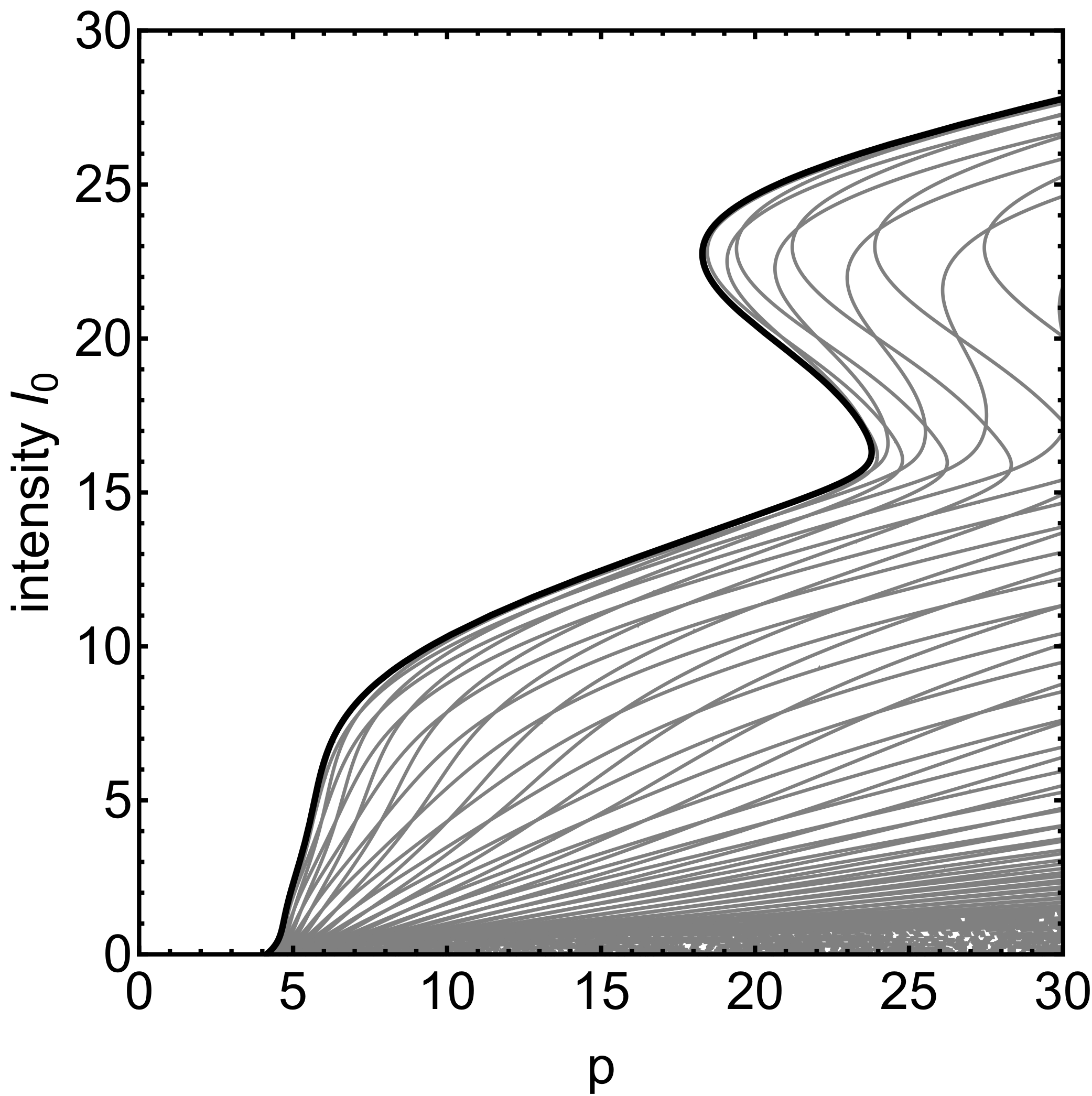}

\caption{\label{fig:ST}CW solutions of the model equations (\ref{eq:Model1})
and (\ref{eq:Model2}) as functions of the pump parameter $p$ (gray
lines). Black line is obtained by substituting $\omega=0$ into Eq.
(\ref{eq:stst1}). Parameters are: $\kappa=0.8$, $a=1.0$, $\alpha=0$,
$T=25$, $K=0.4$, $G=0.5$.}
\end{figure}

Nontrivial continuous wave (CW) solutions of Eqs. (\ref{eq:Model1}) and (\ref{eq:Model2}),
$A\left(t\right)=A_{0}e^{i\omega t}$ and $g=g_{0}$, are defined
by 
\begin{equation}
\frac{\kappa{\cal R}\left(I_{0}\right)e^{g_{0}}}{1+\omega^{2}}=1,\label{eq:stst1}
\end{equation}
\begin{equation}
p-g_{0}-\left(e^{g_{0}}-1\right)I_{0}{\cal R}\left(I_{0}\right)=0.\label{eq:stst2}
\end{equation}
\[
\tan\left(\omega T+\frac{1}{2}\alpha g_{0}\right)=
\]
\begin{equation}
\frac{K\left(\sin Y_{0}-\omega\cos Y_{0}\right)-\left(1-K\right)\left(\sin X_{0}-\omega\cos X_{0}\right)}{K\left(\omega\sin Y_{0}-\cos Y_{0}\right)-\left(1-K\right)\left(\omega\sin X_{0}-\cos X_{0}\right)},\label{eq:stst3}
\end{equation}
where $X_{0}=X\left(I_{0}\right)$ , $Y_{0}=Y\left(I_{0}\right)$
and $I_{0}=|A_{0}|^{2}$. These solutions can be interpreted as longitudinal laser modes. Eq. (\ref{eq:stst1}) can be considered as an energy balance condition, which states that the total losses in the cavity are compensated by the amplification. The intensities $I_{0}$ of different solutions of Eqs. (\ref{eq:stst1}) and (\ref{eq:stst2}) are shown in Fig. \ref{fig:ST} by gray lines as functions of the pump parameter $p$. Black line shows the envelope of these solutions obtained
by substituting $\omega=0$ into Eq. (\ref{eq:stst1}). It is seen
that CW solutions of the the model equations can exhibit a multistable
behavior.

\begin{figure}
\includegraphics[scale=0.45]{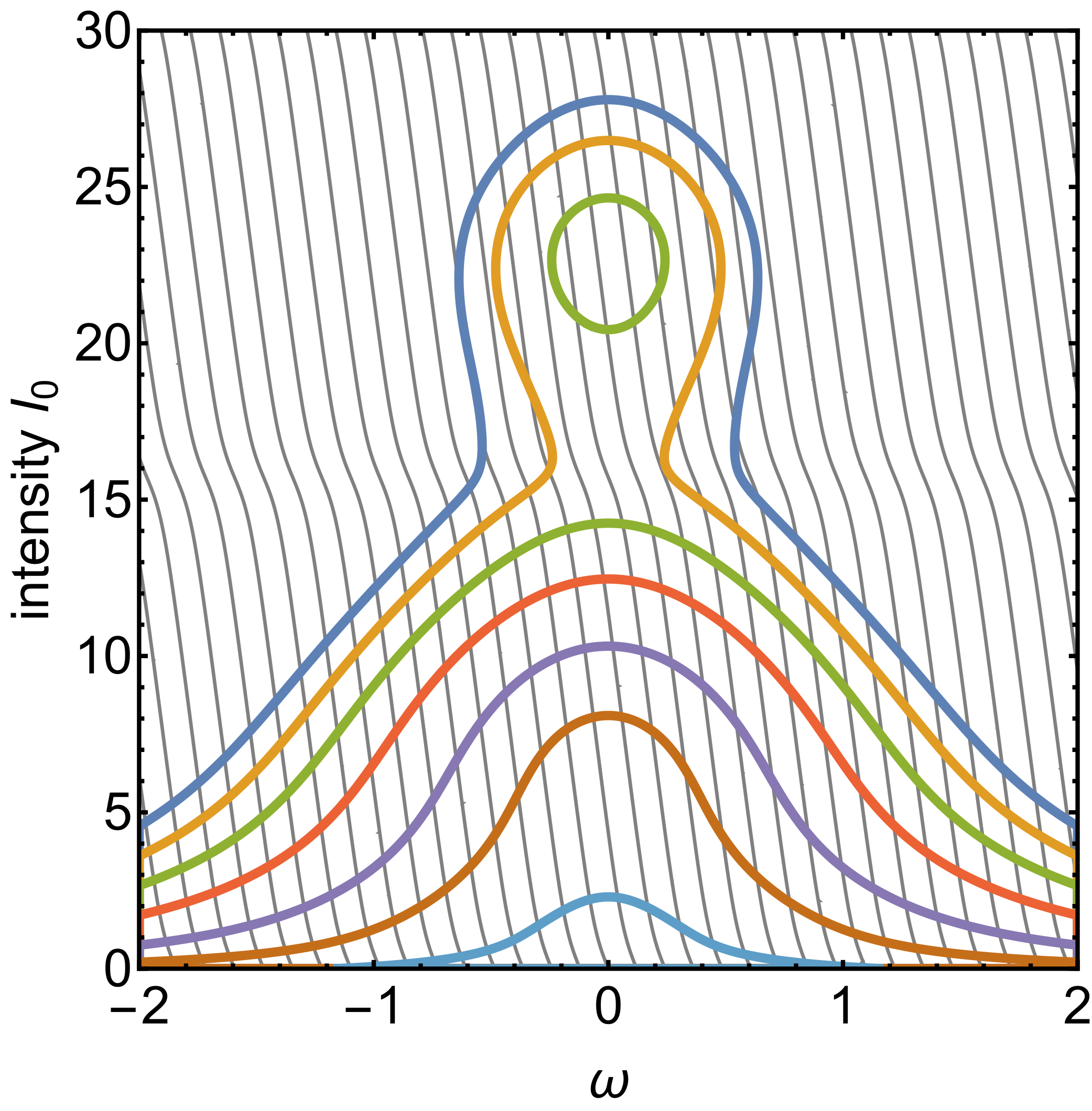}
\caption{\label{fig:CW}CW solutions (longitudinal laser modes) on the ($I_{0}$,$\omega$)-plane
correspond to the intersections of thick colored curves with thin gray lines. Different colors correspond to the pump parameter values: $p=4.0$, $p=4.5$, $p=6.5$, $p=10.0$, $p=13.0$, $p=17.0$, $p=25.0$. Other parameters are the same as in Fig. \ref{fig:ST}.}
\end{figure}

\begin{figure}
\includegraphics[scale=0.45]{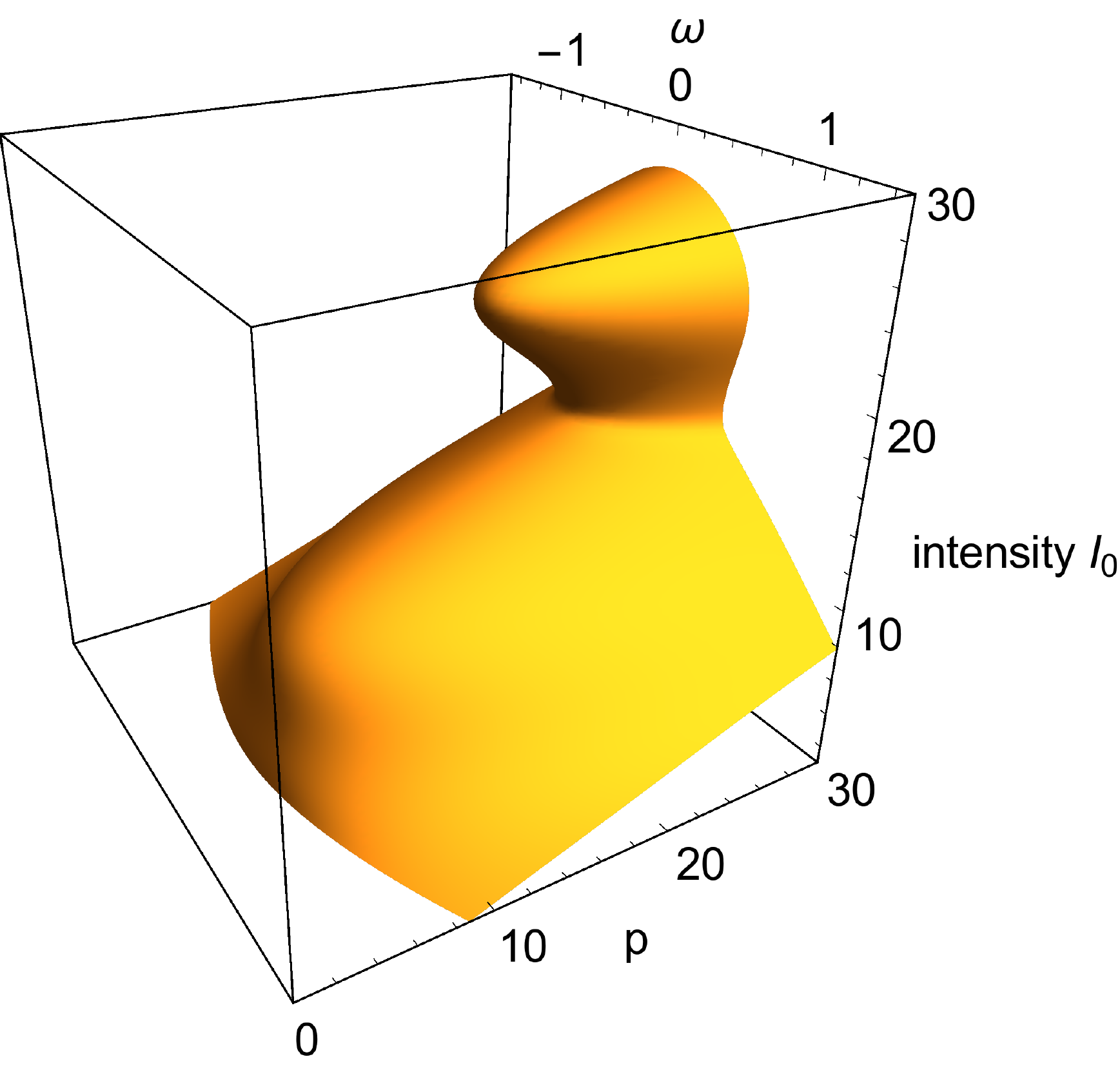}
\caption{\label{fig:Locus}Locus of the CW longitudinal laser modes in the limit $T\to\infty$.
Parameters are the same as in Fig. \ref{fig:ST}. }
\end{figure}

The solutions of Eqs. (\ref{eq:stst1}) and (\ref{eq:stst2}) are
shown in Fig. \ref{fig:CW} for several different values of the pump parameter $p$ by thick colored curves on the ($I_0$,$\omega$)
-plane, while the solutions of Eq. (\ref{eq:stst3}) are indicated
by thin gray lines. The intersections of thick colored lines with thin gray lines correspond to CW longitudinal laser modes. The density of gray lines increases with $T$, so that in the limit $T\to\infty$ the modes fill densely the colored curves. Therefore, in this limit thick colored curves in Fig. \ref{fig:CW} defined by Eqs. (\ref{eq:stst1})
and (\ref{eq:stst2}) determine the locus of the CW laser modes. A
representation of this locus as a 2D surface in the 3D space $(p,\omega,I_{0})$
is shown in Fig. \ref{fig:Locus}. 

\section{Linear stability analysis in the limit of large delay}

\begin{figure}
\includegraphics[scale=0.45]{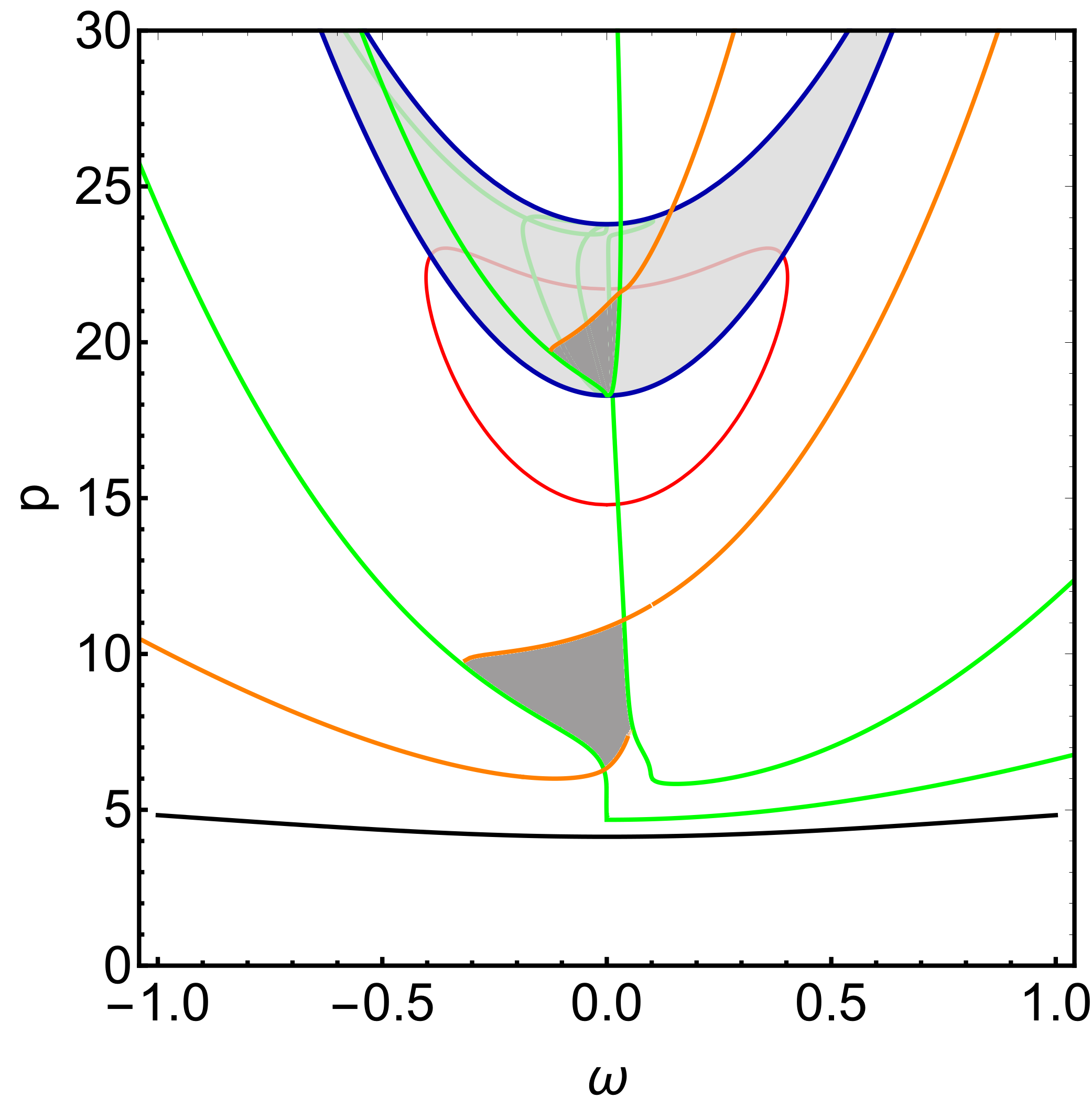}
\caption{\label{fig:diagram}Bifurcation diagram of the CW solutions in the
limit $T\to\infty$. Light gray area shows bistability domain where the solutions of Eqs. (\ref{eq:stst1}) and (\ref{eq:stst2}) have three nontrivial solutions for $I_0>0$. This area is limited by the saddle-node instability boundary indicated by blue line. Dark gray areas indicate the stability domains of CW regimes. Green line indicates long wavelength modulation instability, orange
line -- short wavelength modulation instability, and red line --
flip instability. $\gamma=0.05$. Other parameters are the same as
in Fig. \ref{fig:ST}.}

\end{figure}

In the large delay limit $T\to\infty$ we assume that the CW solutions of Eqs. (\ref{eq:Model1}) and (\ref{eq:Model2}) fill densely the locus of the CW solutions defined by Eqs. (\ref{eq:stst1}) and (\ref{eq:stst2}).
In this case we can forget about Eq. (\ref{eq:stst3}) and consider
the frequency $\omega$ of the CW solutions as a pseudo-continuous
variable. The bifurcation diagram calculated in the limit
$T\to\infty$ is shown in Fig. \ref{fig:diagram}. In this figure linear laser threshold defined by the condition 
\[
\frac{\kappa{\cal R}\left(0\right)e^{p}}{1+\omega^{2}}=1,
\]
is indicated by black line on the $\left(\omega,p\right)$-plane. 

Linear stability of a nontrivial CW solution with the frequency $\omega$ is determined by the solutons $\lambda$ of the characteristic equation
\begin{equation}
c_{2}\left(\lambda\right)Y^{2}+c_{1}\left(\lambda\right)Y+c_{0}\left(\lambda\right)=0,\label{eq:characteristic}
\end{equation}
where 
\[
Y=e^{-\lambda T},
\]
and the expressions for the coefficients $c_{0}\left(\lambda\right)$,
$c_{1}\left(\lambda\right)$, and $c_{2}\left(\lambda\right)$ are
given in the Appendix \ref{Appendix}.

In the limit $T\to\infty$ the pseudo-continuous spectrum $\mu\left(\nu\right)$of the CW solutions
is obtained by solving characteristic equation (\ref{eq:characteristic})
with respect to $Y$, $Y=Y_{1,2}\left(\lambda\right)$, and performing the substitution $\lambda\to i\nu$ in the resulting solution \cite{Yanchuk2010a}:
\[
\mu\left(\nu\right)=-\ln\left[Y\left(i\nu\right)\right].
\]

Saddle-node and flip instabilities of CW solutions are defined by
the condition that the first solution of the characteristic equation
(\ref{eq:characteristic}) satisfies the conditions
\begin{equation}
Y_{1}\left(0\right)=1\label{eq:SN}
\end{equation}
and
\begin{equation}
Y_{1}\left(0\right)=-1,\label{eq:flip}
\end{equation}
respectively, with
\begin{equation}
Y_{1}\left(0\right)=\frac{\kappa+I_{0}\left(1+\omega^{2}\right)}{\kappa\left[1+I_{0}{\cal R}\left(I_{0}\right)\right]\left[1+I_{0}\frac{d\ln{\cal R}\left(I_{0}\right)}{dI_{0}}\right]}.\label{eq:Y1(0)}
\end{equation}
It follows from Eq. (\ref{eq:Y1(0)}) that saddle-node and flip instability conditions do not depend on the carrier relaxation rate $\gamma$. The saddle node instability condition (\ref{eq:SN}) defines the folds of CW locus surface shown in Fig. \ref{fig:Locus}. 
The flip instability (\ref{eq:flip}) is responsible for a period-doubling cascade giving rise to more and more complicated square wave patterns with increasing periods \cite{vladimirov2019dynamics}. 
Light gray area in Fig. \ref{fig:diagram} indicates the bistability domain where for every given frequency $\omega$ there are three nontrivial solutions of Eqs. (\ref{eq:stst1}) and (\ref{eq:stst2}) with $I_{0}>0$. This domain is limited by the saddle-node instability boundary shown by blue line.  Red line indicates the flip instability.

Unlike the flip and saddle-node instabilities, short and long wavelength modulational instabilities of the CW solutions depend on the normalized inversion relaxation rate $\gamma$. The second solution $Y_{2}\left(\lambda\right)$ of the characteristic
equation (\ref{eq:characteristic}) has the property $Y_{2}\left(0\right)=1$ or equivalently $\mu_{2}\left(0\right)=0$, which corresponds to the phase shift symmetry of the model equations (\ref{eq:Model1}) and
(\ref{eq:Model2}), $A\left(t\right)\to A\left(t\right)e^{i\phi}$
with arbitrary $\phi$. The long wavelength modulational instability is defined by the condition
\[
\Re\left[\partial_{\nu\nu}\mu_{2}\left(\nu\right)\right]_{\nu=0}=\Re\left\{ \partial_{\nu\nu}\left[-\ln Y_{2}\left(i\nu\right)\right]\right\} _{\nu=0}=0.
\]
It is shown by green lines in Fig. \ref{fig:diagram}. Another type
of the modulational instability of CW solutions of Eqs. (\ref{eq:Model1})
and (\ref{eq:Model2}) is the short wavelength instability which corresponds
to the situation when the pseudo-continuous spectral curve touches
the imaginary axis at the points $\nu=\pm\nu_{0}$ with $\nu_{0}>0$,
i.e.
\[
\Re\left[\mu_{1}\left(\nu_{0}\right)\right]=\Re\left[\partial_{\nu}\mu_{1}\left(\nu\right)\right]_{\nu=\nu_{0}}=0,
\]
where $\mu_{1}=-\ln Y_{1}\left(i\nu\right)$. This instability is
shown in Fig. \ref{fig:diagram} by orange lines. It is seen that unlike the case of adiabatically eliminated population inversion when stable CW solutions can exhibit a flip instability and period doubling cascade leading to a formation of square wave patterns \cite{vladimirov2019dynamics},
for the parameters of Fig. \ref{fig:diagram} corresponding to a relatively slow inversion relaxation short wavelength modulational instability takes place before the flip instability. Therefore, unlike the case of adiabatically eliminated gain variable studied in Ref. \cite{vladimirov2019dynamics}, bifurcation
of stable square wave patterns from CW solutions via a cascade of
flip instabilities is not possible for such parameter values.

\section{Numerical results}

\begin{figure}
\includegraphics[scale=0.45]{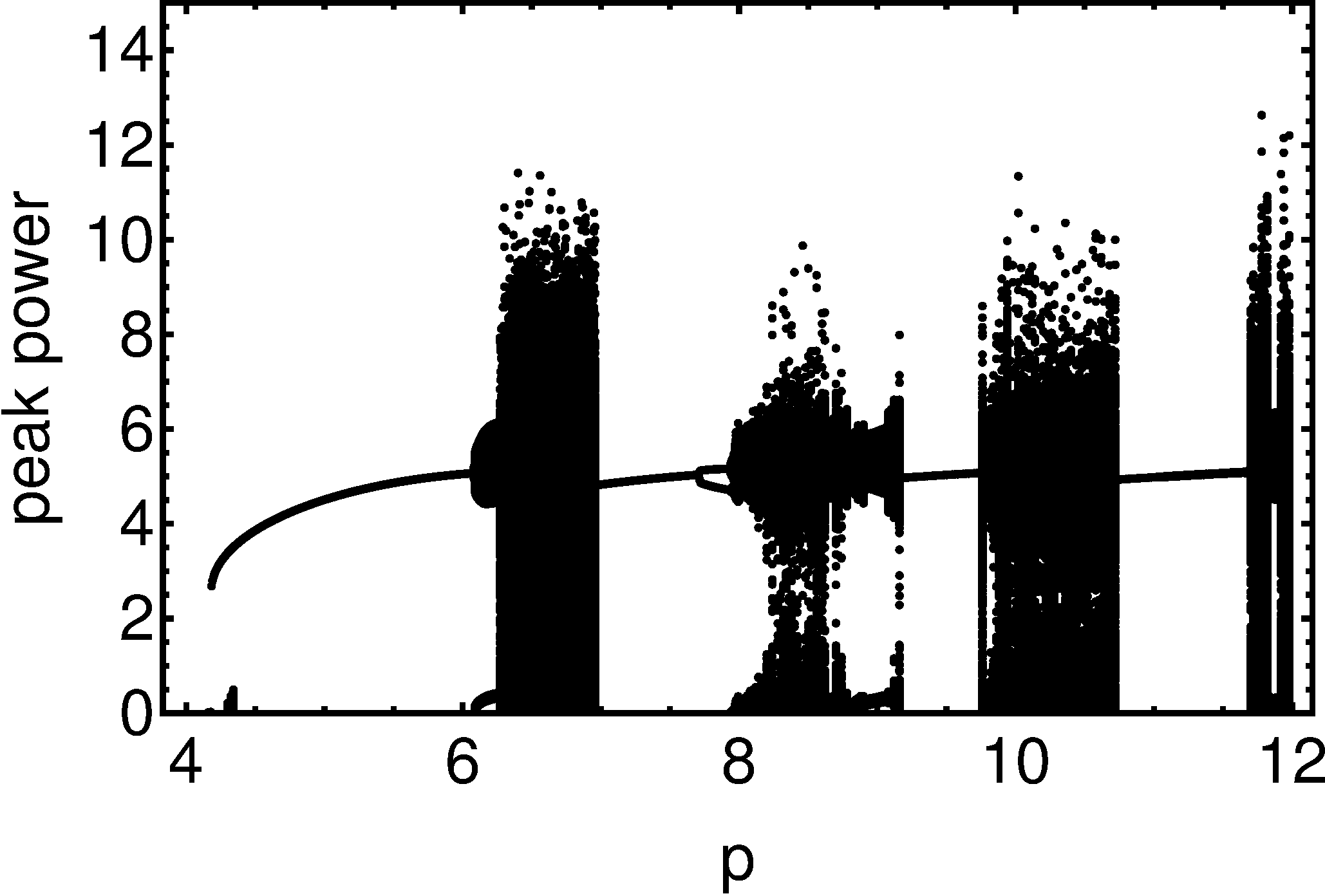}3

\caption{\label{fig:Biftree}Bifurcation tree illustrating the pulse peak power as function of the pump parameter
$p$. $\kappa=0.8$, $a=1.0$, $\alpha=0$, $T=25$, $K=0.4$, $G=0.5$. }

\end{figure}

\begin{figure}
\includegraphics[scale=0.4]{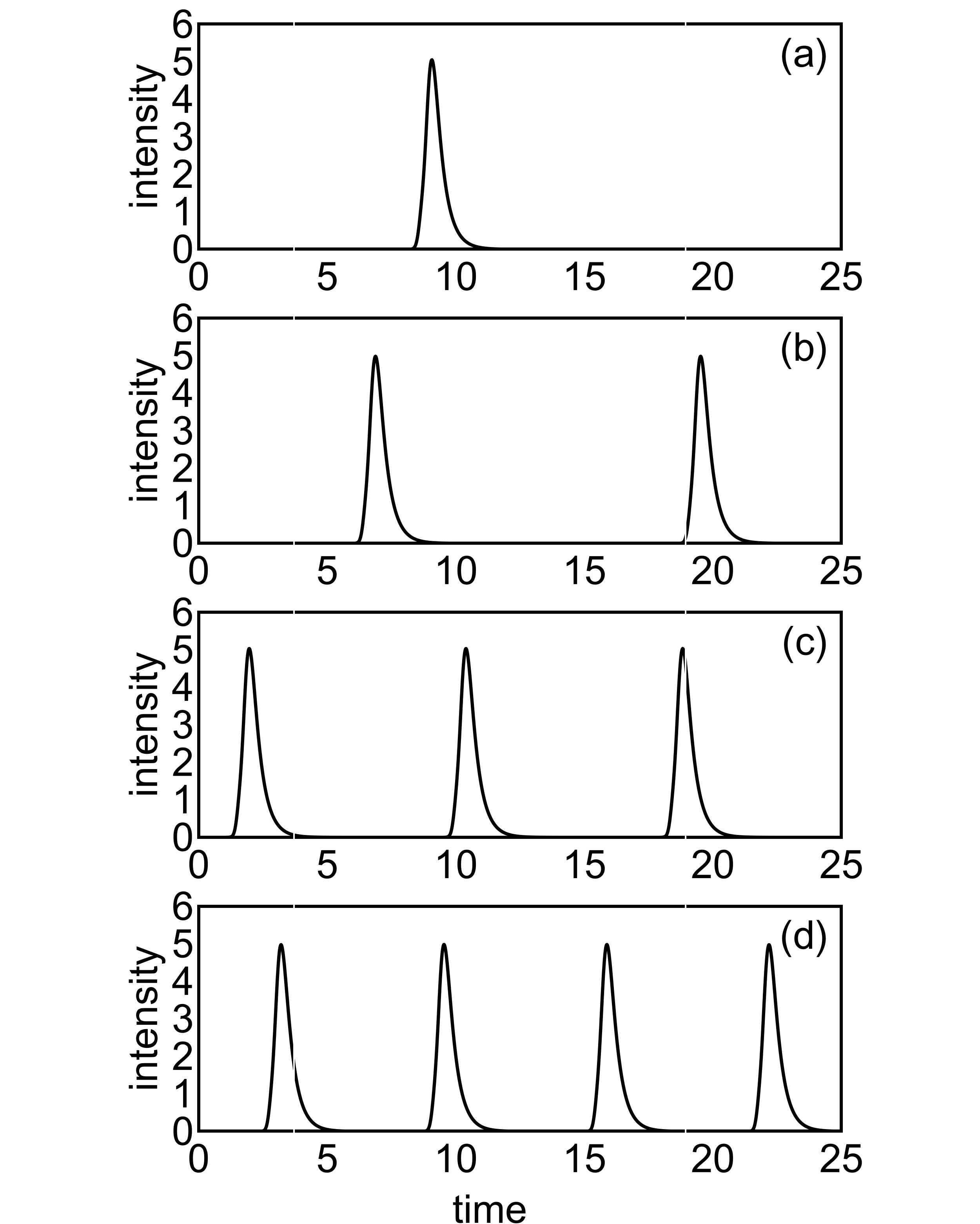}

\caption{\label{fig:timetraces}Fundamental mode-locking regime (a), $p=6.0$. Harmonic mode-locking regimes with two (b), $p=7.5$; three (c), $p=9.5$; and
four (d), $p=11.5$, pulses per cavity round trip time. Other parameters
are the same as in Fig. \ref{fig:Biftree}.}
\end{figure}

Bifurcation tree obtained by numerical integration of Eqs. (\ref{eq:Model1})-(\ref{eq:Model2})
is shown in Fig. \ref{fig:Biftree}. Black dots in this figure correspond
to local maximums of the laser field intensity time trace (pulse peak powers) calculated after the transient time of $4000$ cavity round trips. For a given pump parameter value where the laser exhibits a regular mode-locked regime all the dots coincide and have their ordinate equal to the pulse peak power. Irregular pulsing behavior corresponds to a cloud of dots having different ordinates corresponding to different pulse peak powers.
The bifurcation tree in Fig. \ref{fig:Biftree} shows four windows of regular mode-locking regimes separated by the domains of irregular pulsing. The first, second, third, and fourth window correspond to a regime with one, two, three, and four pulses per cavity round trip, respectively. Fundamental mode-locking regime is illustrated in Fig. \ref{fig:timetraces}(a), while harmonic mode-locked regimes with two, three, and four pulses per cavity round trip time are shown in Figs. \ref{fig:timetraces}(b),
(c), and (d), respectively. It is seen that although initially the pulse peak power of the fundamental mode-locking regime grows with the pump parameter $p$, further increase of $p$ leads to an increase of the number of pulses per cavity round trip time, while pulse peak power and shape remain almost unchanged. The pulses shown in Fig.
\ref{fig:timetraces} are asymmetric with slowly decaying trailing
edge due to relatively slow relaxation of the carrier density in the gain medium. Note that unlike the case of 50:50 beam splitter  \cite{vladimirov2019dynamics}, where mode-locked pulses are always bistable with the laser off state, for the parameter values of Fig. \ref{fig:Biftree} corresponding to asymmetric 40:60 splitter stable mode-locked pulses exist above the linear laser threshold, where the laser off solution is unstable. 

\begin{figure}
\includegraphics[scale=0.4]{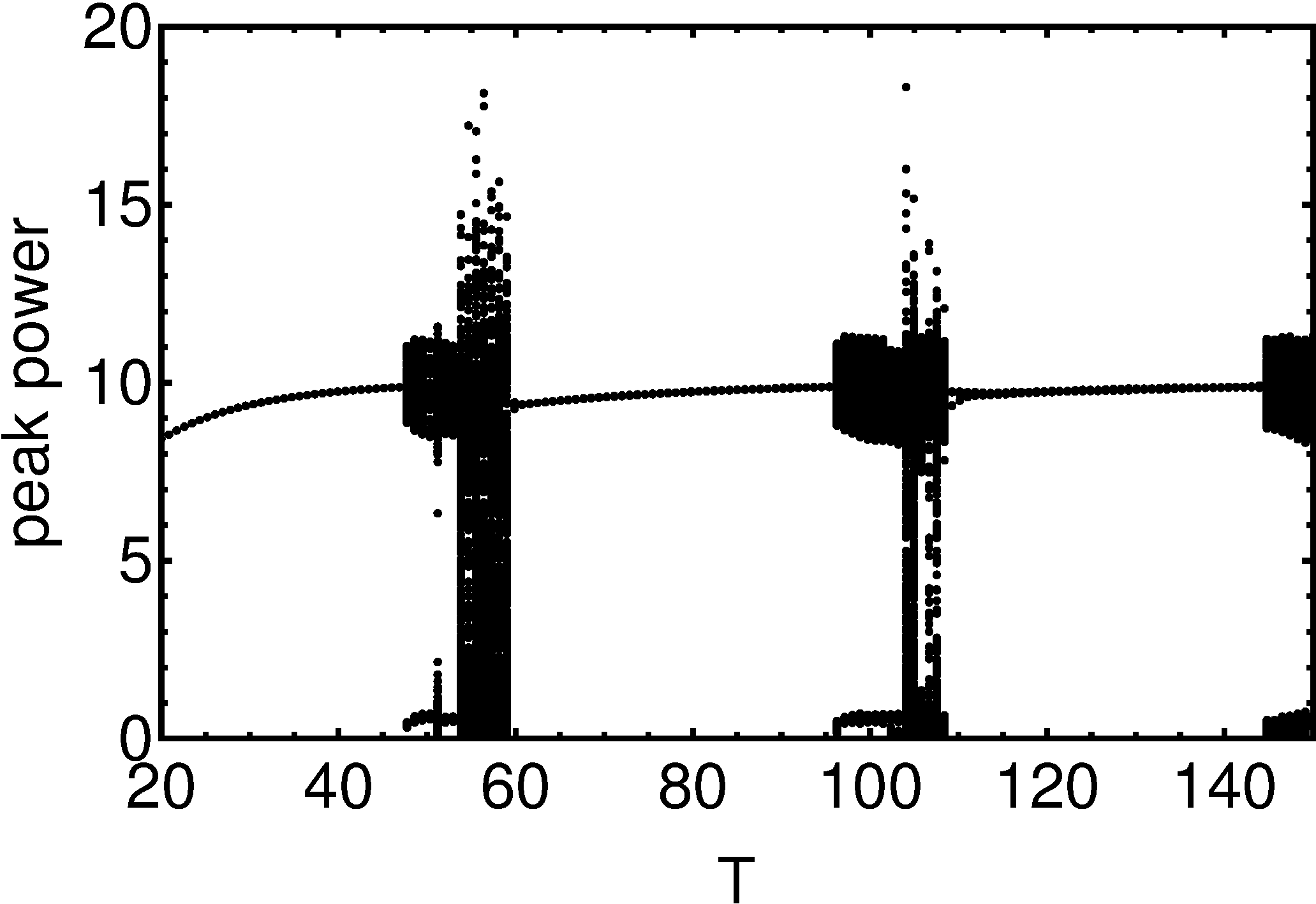}

\caption{Bifurcation tree similar to that shown in Fig. \ref{fig:Biftree},
but obtained by changing the cold cavity round trip time $T$. $p=5$.
Other parameters are the same as in Fig. \ref{fig:Biftree}\label{fig:Bif-T}}
\end{figure}
In Fig. \ref{fig:Bif-T} presenting a bifurcation tree similar
to that shown in Fig. \ref{fig:Biftree}, but obtained by increasing the delay parameter $T$, three windows of regular mode-locking solutions are separated by thin domains of irregular pulsing. Here the first, second, and third mode-locking window correspond to regular mode-locking regimes with one, two, and three pulses per cavity round trip time.
These regimes are similar to those shown in Fig. \ref{fig:timetraces}(a),(b),
and (c).

\begin{figure}
    \centering
    \includegraphics[width=\columnwidth]{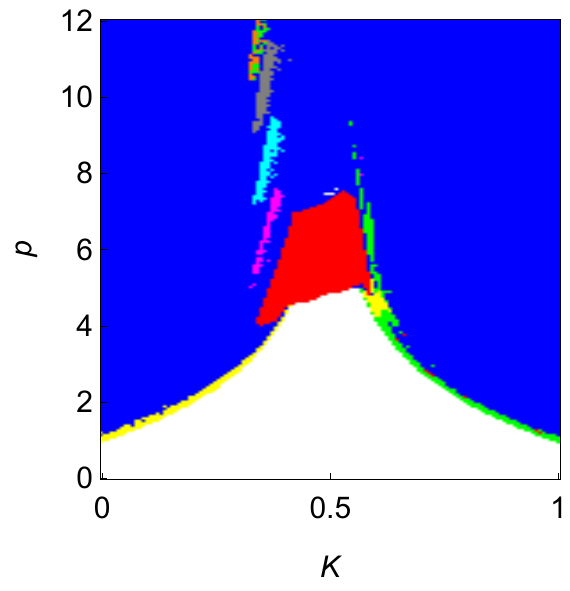}
    \caption{Map of different dynamical regimes in the ($K$,$p$)-plane.  White color indicates laser off regime,  yellow - CW solutions, and green -- weakly periodically modulated CW solutions. Red area corresponds to fundamental mode-locked regime with a single pulse per cavity round trip. Magenta, cyan and gray -- to harmonic mode-locked regimes with two, three, and four pulses per cavity round trip. Blue color indicates irregular pulsing, Parameters are the same as in Fig.~\ref{fig:ST}}
    \label{FigReg}
\end{figure}

Our numerical simulation show that for certain parameter values the solutions of the model equations can exhibit bistability or multistability when choosing different initial conditions. In particular, it follows from Figs. \ref{fig:diagram} and \ref{fig:Biftree} that stable mode-locked pulses can coexist with stable CW regimes. In addition, these pulses can coexist with irregular pulsed regimes and, when $K$ is sufficiently close to $0.5$, with laser off state.
Numerically calculated map of dynamical regimes is shown in Fig.~\ref{FigReg} in the two-parameter plane ($K$,$p$). It was obtained by integration of  Eq.~\eqref{eq:Model1} and \eqref{eq:Model2} with the initial condition in the form of Gaussian pulse: $A(t)=A_m\exp[-(t+0.5 T)^2/w^2]$ and $g(t)=p$ on the interval $t\in[-T,0]$, where $A_m=2$ and $w=4$. Red color in Fig.~\ref{FigReg} indicates the region of fundamental mode-locked regime with a single pulse per cavity round trip. Regions of harmonic mode-locked regimes with two, three, and four pulses per cavity round trip are shown by magenta, cyan, and gray colors, respectively. White color corresponds to the laser off regime. Yellow and green colors indicate stable and weakly periodically modulated CW regimes, respectively. It is seen from Fig.~\ref{FigReg} that the domain of the fundamental mode-locked regime is asymmetric in $K$ and located around $K=0.5$ and that the use of slightly asymmetric beam splitter could help to achieve stable harmonic mode-locked regimes. 

\section{Conclusion}

We have developed and analyzed a DDE  NOLM-NALM mode-locked
laser model taking into account arbitrary inversion relaxation rate in the gain medium as well as asymmetry of the beam splitter. Our numerical simulations indicate that with increasing pump parameter this model can exhibit large windows of regular fundamental and harmonic mode-locked regimes separated by regions of irregular pulsing. Experimental observation harmonic mode-locking regimes in NOLM-NALM lasers was reported in e.g. in \cite{zhang2008passively,lee2015femtosecond,deng2020energy}. We
have shown that unlike the laser with symmetric beam splitter where
mode-locked pulses always coexist with a stable laser off solution
a laser with asymmetric beam splitter can exhibit regular mode-locked regimes above the linear lasing threshold where the laser off solution is unstable. Our numerical simulations indicate that a proper choice of the beam splitting ration can favor the development of harmonic mode-locked regimes. Furthermore, we have demonstrated that sufficiently slow relaxation of the gain inversion can lead to a suppression of the flip instability leading to a period
doubling bifurcation cascade and the formation of square wave patterns in the laser output. This instability was predicted theoretically using a Poincare map NOLM-NALM laser model \cite{lai2005nolm} as well as DDE models with adiabatically eliminated gain \cite{vladimirov2019dynamics,aadhi2019highly}.
Experimental observation of square wave patterns was reported in \cite{aadhi2019highly}, in a NALM laser with SOA amplifier in the nonlinear mirror loop. On the other hand experimental studies of \cite{lai2005nolm} have not revealed period doubling cascade predicted theoretically in the same paper. We believe that this work could create a theoretical basis for further steps in modeling of specific types of lasers, for instance, including into consideration the effect of chromatic dispersion of the intracavity media. 
\begin{acknowledgments}
We gratefully acknowledge the suport by the Deutsche Forschungsgemeinschaft
(DFG-RSF project No.445430311). Work of S.S. and S.K.T. has been supported by the Russian Science Foundation (RSF-DFG project  21-42-04401)
\end{acknowledgments}

\appendix
\section{Coefficients of the characteristic equation (\ref{eq:characteristic})}\label{Appendix}
The coefficients $c_{0,1,2}\left(\lambda\right)$ in the characteristic
equations \ref{eq:characteristic} are given by:
\begin{widetext}
\[
c_{0}\left(\lambda\right)=\left[\left(1+\lambda\right)^{2}+\omega^{2}\right]\left[\gamma+\frac{\gamma I_{0}}{\kappa}\left(1+\omega^{2}\right)+\lambda\right],
\]
\[
c_{1}=-\left(1+\lambda+\omega^{2}\right)\left\{ 2\left(\gamma+\lambda\right)+\frac{\gamma I_{0}}{\kappa}\left[1+\omega^{2}+\kappa{\cal R}\left(I_{0}\right)\right]+I_{0}\left[\text{\ensuremath{\gamma}}+\text{\ensuremath{\lambda+\gamma}}I_{0}{\cal R}\left(I_{0}\right)\right]\frac{d\ln{\cal R}\left(I_{0}\right)}{dI_{0}}\right\} -\lambda\omega W\left[\gamma+\lambda+\frac{\gamma I_{0}}{\kappa}\left(1+\omega^{2}\right)\right],
\]
\[
c_{2}\left(\lambda\right)=\left(1+\omega^{2}\right)\left[\gamma+\gamma I_{0}{\cal R}\left(I_{0}\right)+\lambda\right]\left[1+I_{0}\frac{d\ln{\cal R}\left(I_{0}\right)}{dI_{0}}\right],
\]
\end{widetext}
where $W=aI_{0}\left[\frac{\left(1-2K\right)\left(1-K-GK\right)}{{\cal R}\left(I_{0}\right)}+1-\left(1-G\right)K\right]$
and $I_{0}=|A_{0}|^{2}$ is the CW intensity obtained from the solution
of Eqs. (\ref{eq:stst1}) and (\ref{eq:stst2}).

\bibliographystyle{apsrev4-2}

\begin{thebibliography}{30}%
\makeatletter
\providecommand \@ifxundefined [1]{%
 \@ifx{#1\undefined}
}%
\providecommand \@ifnum [1]{%
 \ifnum #1\expandafter \@firstoftwo
 \else \expandafter \@secondoftwo
 \fi
}%
\providecommand \@ifx [1]{%
 \ifx #1\expandafter \@firstoftwo
 \else \expandafter \@secondoftwo
 \fi
}%
\providecommand \natexlab [1]{#1}%
\providecommand \enquote  [1]{``#1''}%
\providecommand \bibnamefont  [1]{#1}%
\providecommand \bibfnamefont [1]{#1}%
\providecommand \citenamefont [1]{#1}%
\providecommand \href@noop [0]{\@secondoftwo}%
\providecommand \href [0]{\begingroup \@sanitize@url \@href}%
\providecommand \@href[1]{\@@startlink{#1}\@@href}%
\providecommand \@@href[1]{\endgroup#1\@@endlink}%
\providecommand \@sanitize@url [0]{\catcode `\\12\catcode `\$12\catcode
  `\&12\catcode `\#12\catcode `\^12\catcode `\_12\catcode `\%12\relax}%
\providecommand \@@startlink[1]{}%
\providecommand \@@endlink[0]{}%
\providecommand \url  [0]{\begingroup\@sanitize@url \@url }%
\providecommand \@url [1]{\endgroup\@href {#1}{\urlprefix }}%
\providecommand \urlprefix  [0]{URL }%
\providecommand \Eprint [0]{\href }%
\providecommand \doibase [0]{https://doi.org/}%
\providecommand \selectlanguage [0]{\@gobble}%
\providecommand \bibinfo  [0]{\@secondoftwo}%
\providecommand \bibfield  [0]{\@secondoftwo}%
\providecommand \translation [1]{[#1]}%
\providecommand \BibitemOpen [0]{}%
\providecommand \bibitemStop [0]{}%
\providecommand \bibitemNoStop [0]{.\EOS\space}%
\providecommand \EOS [0]{\spacefactor3000\relax}%
\providecommand \BibitemShut  [1]{\csname bibitem#1\endcsname}%
\let\auto@bib@innerbib\@empty
\bibitem [{\citenamefont {Doran}\ and\ \citenamefont
  {Wood}(1988)}]{doran1988nonlinear}%
  \BibitemOpen
  \bibfield  {author} {\bibinfo {author} {\bibfnamefont {N.}~\bibnamefont
  {Doran}}\ and\ \bibinfo {author} {\bibfnamefont {D.}~\bibnamefont {Wood}},\
  }\href@noop {} {\bibfield  {journal} {\bibinfo  {journal} {Opt. Lett.}\
  }\textbf {\bibinfo {volume} {13}},\ \bibinfo {pages} {56} (\bibinfo {year}
  {1988})}\BibitemShut {NoStop}%
\bibitem [{\citenamefont {Richardson}\ \emph {et~al.}(1991)\citenamefont
  {Richardson}, \citenamefont {Laming}, \citenamefont {Payne}, \citenamefont
  {Matsas},\ and\ \citenamefont {Phillips}}]{F801}%
  \BibitemOpen
  \bibfield  {author} {\bibinfo {author} {\bibfnamefont {D.}~\bibnamefont
  {Richardson}}, \bibinfo {author} {\bibfnamefont {R.}~\bibnamefont {Laming}},
  \bibinfo {author} {\bibfnamefont {D.}~\bibnamefont {Payne}}, \bibinfo
  {author} {\bibfnamefont {V.}~\bibnamefont {Matsas}},\ and\ \bibinfo {author}
  {\bibfnamefont {M.}~\bibnamefont {Phillips}},\ }\href@noop {} {\bibfield
  {journal} {\bibinfo  {journal} {Electronics Letters}\ }\textbf {\bibinfo
  {volume} {27}},\ \bibinfo {pages} {542} (\bibinfo {year} {1991})}\BibitemShut
  {NoStop}%
\bibitem [{\citenamefont {Duling}(1991)}]{F802}%
  \BibitemOpen
  \bibfield  {author} {\bibinfo {author} {\bibfnamefont {I.}~\bibnamefont
  {Duling}},\ }\href
  {https://digital-library.theiet.org/content/journals/10.1049/el_19910342}
  {\bibfield  {journal} {\bibinfo  {journal} {Electronics Letters}\ }\textbf
  {\bibinfo {volume} {27}},\ \bibinfo {pages} {544} (\bibinfo {year}
  {1991})}\BibitemShut {NoStop}%
\bibitem [{\citenamefont {Theimer}\ and\ \citenamefont
  {Haus}(1997)}]{theimer1997figure}%
  \BibitemOpen
  \bibfield  {author} {\bibinfo {author} {\bibfnamefont {J.}~\bibnamefont
  {Theimer}}\ and\ \bibinfo {author} {\bibfnamefont {J.}~\bibnamefont {Haus}},\
  }\href@noop {} {\bibfield  {journal} {\bibinfo  {journal} {Journal of modern
  Optics}\ }\textbf {\bibinfo {volume} {44}},\ \bibinfo {pages} {919} (\bibinfo
  {year} {1997})}\BibitemShut {NoStop}%
\bibitem [{\citenamefont {Salhi}\ \emph {et~al.}(2008)\citenamefont {Salhi},
  \citenamefont {Haboucha}, \citenamefont {Leblond},\ and\ \citenamefont
  {Sanchez}}]{salhi2008theoretical}%
  \BibitemOpen
  \bibfield  {author} {\bibinfo {author} {\bibfnamefont {M.}~\bibnamefont
  {Salhi}}, \bibinfo {author} {\bibfnamefont {A.}~\bibnamefont {Haboucha}},
  \bibinfo {author} {\bibfnamefont {H.}~\bibnamefont {Leblond}},\ and\ \bibinfo
  {author} {\bibfnamefont {F.}~\bibnamefont {Sanchez}},\ }\href@noop {}
  {\bibfield  {journal} {\bibinfo  {journal} {Physical Review A}\ }\textbf
  {\bibinfo {volume} {77}},\ \bibinfo {pages} {033828} (\bibinfo {year}
  {2008})}\BibitemShut {NoStop}%
\bibitem [{\citenamefont {Li}\ \emph {et~al.}(2016)\citenamefont {Li},
  \citenamefont {Li}, \citenamefont {Zhou}, \citenamefont {Zhao}, \citenamefont
  {Tang},\ and\ \citenamefont {Shen}}]{li2016characterization}%
  \BibitemOpen
  \bibfield  {author} {\bibinfo {author} {\bibfnamefont {D.}~\bibnamefont
  {Li}}, \bibinfo {author} {\bibfnamefont {L.}~\bibnamefont {Li}}, \bibinfo
  {author} {\bibfnamefont {J.}~\bibnamefont {Zhou}}, \bibinfo {author}
  {\bibfnamefont {L.}~\bibnamefont {Zhao}}, \bibinfo {author} {\bibfnamefont
  {D.}~\bibnamefont {Tang}},\ and\ \bibinfo {author} {\bibfnamefont
  {D.}~\bibnamefont {Shen}},\ }\href@noop {} {\bibfield  {journal} {\bibinfo
  {journal} {Scientific reports}\ }\textbf {\bibinfo {volume} {6}},\ \bibinfo
  {pages} {1} (\bibinfo {year} {2016})}\BibitemShut {NoStop}%
\bibitem [{\citenamefont {Smirnov}\ \emph {et~al.}(2017)\citenamefont
  {Smirnov}, \citenamefont {Kobtsev}, \citenamefont {Ivanenko}, \citenamefont
  {Kokhanovskiy}, \citenamefont {Kemmer},\ and\ \citenamefont
  {Gervaziev}}]{smirnov2017layout}%
  \BibitemOpen
  \bibfield  {author} {\bibinfo {author} {\bibfnamefont {S.}~\bibnamefont
  {Smirnov}}, \bibinfo {author} {\bibfnamefont {S.}~\bibnamefont {Kobtsev}},
  \bibinfo {author} {\bibfnamefont {A.}~\bibnamefont {Ivanenko}}, \bibinfo
  {author} {\bibfnamefont {A.}~\bibnamefont {Kokhanovskiy}}, \bibinfo {author}
  {\bibfnamefont {A.}~\bibnamefont {Kemmer}},\ and\ \bibinfo {author}
  {\bibfnamefont {M.}~\bibnamefont {Gervaziev}},\ }\href@noop {} {\bibfield
  {journal} {\bibinfo  {journal} {Optics letters}\ }\textbf {\bibinfo {volume}
  {42}},\ \bibinfo {pages} {1732} (\bibinfo {year} {2017})}\BibitemShut
  {NoStop}%
\bibitem [{\citenamefont {Cai}\ \emph {et~al.}(2017)\citenamefont {Cai},
  \citenamefont {Chen}, \citenamefont {Chen},\ and\ \citenamefont
  {Hou}}]{cai2017state}%
  \BibitemOpen
  \bibfield  {author} {\bibinfo {author} {\bibfnamefont {J.-H.}\ \bibnamefont
  {Cai}}, \bibinfo {author} {\bibfnamefont {H.}~\bibnamefont {Chen}}, \bibinfo
  {author} {\bibfnamefont {S.-P.}\ \bibnamefont {Chen}},\ and\ \bibinfo
  {author} {\bibfnamefont {J.}~\bibnamefont {Hou}},\ }\href@noop {} {\bibfield
  {journal} {\bibinfo  {journal} {Optics express}\ }\textbf {\bibinfo {volume}
  {25}},\ \bibinfo {pages} {4414} (\bibinfo {year} {2017})}\BibitemShut
  {NoStop}%
\bibitem [{\citenamefont {Boscolo}\ \emph {et~al.}(2019)\citenamefont
  {Boscolo}, \citenamefont {Finot}, \citenamefont {Gukov},\ and\ \citenamefont
  {Turitsyn}}]{boscolo2019performance}%
  \BibitemOpen
  \bibfield  {author} {\bibinfo {author} {\bibfnamefont {S.}~\bibnamefont
  {Boscolo}}, \bibinfo {author} {\bibfnamefont {C.}~\bibnamefont {Finot}},
  \bibinfo {author} {\bibfnamefont {I.}~\bibnamefont {Gukov}},\ and\ \bibinfo
  {author} {\bibfnamefont {S.~K.}\ \bibnamefont {Turitsyn}},\ }\href@noop {}
  {\bibfield  {journal} {\bibinfo  {journal} {Laser Physics Letters}\ }\textbf
  {\bibinfo {volume} {16}},\ \bibinfo {pages} {065105} (\bibinfo {year}
  {2019})}\BibitemShut {NoStop}%
\bibitem [{\citenamefont {Deng}\ \emph {et~al.}(2020)\citenamefont {Deng},
  \citenamefont {Zhang}, \citenamefont {Gong}, \citenamefont {He},
  \citenamefont {Li},\ and\ \citenamefont {Gong}}]{deng2020energy}%
  \BibitemOpen
  \bibfield  {author} {\bibinfo {author} {\bibfnamefont {D.}~\bibnamefont
  {Deng}}, \bibinfo {author} {\bibfnamefont {H.}~\bibnamefont {Zhang}},
  \bibinfo {author} {\bibfnamefont {Q.}~\bibnamefont {Gong}}, \bibinfo {author}
  {\bibfnamefont {L.}~\bibnamefont {He}}, \bibinfo {author} {\bibfnamefont
  {D.}~\bibnamefont {Li}},\ and\ \bibinfo {author} {\bibfnamefont
  {M.}~\bibnamefont {Gong}},\ }\href@noop {} {\bibfield  {journal} {\bibinfo
  {journal} {Optics \& Laser Technology}\ }\textbf {\bibinfo {volume} {125}},\
  \bibinfo {pages} {106010} (\bibinfo {year} {2020})}\BibitemShut {NoStop}%
\bibitem [{\citenamefont {Vladimirov}\ \emph {et~al.}(2019)\citenamefont
  {Vladimirov}, \citenamefont {Kovalev}, \citenamefont {Viktorov},
  \citenamefont {Rebrova},\ and\ \citenamefont
  {Huyet}}]{vladimirov2019dynamics}%
  \BibitemOpen
  \bibfield  {author} {\bibinfo {author} {\bibfnamefont {A.~G.}\ \bibnamefont
  {Vladimirov}}, \bibinfo {author} {\bibfnamefont {A.~V.}\ \bibnamefont
  {Kovalev}}, \bibinfo {author} {\bibfnamefont {E.~A.}\ \bibnamefont
  {Viktorov}}, \bibinfo {author} {\bibfnamefont {N.}~\bibnamefont {Rebrova}},\
  and\ \bibinfo {author} {\bibfnamefont {G.}~\bibnamefont {Huyet}},\
  }\href@noop {} {\bibfield  {journal} {\bibinfo  {journal} {Physical Review
  E}\ }\textbf {\bibinfo {volume} {100}},\ \bibinfo {pages} {012216} (\bibinfo
  {year} {2019})}\BibitemShut {NoStop}%
\bibitem [{\citenamefont {Vladimirov}\ \emph
  {et~al.}(2004{\natexlab{a}})\citenamefont {Vladimirov}, \citenamefont
  {Turaev},\ and\ \citenamefont {Kozyreff}}]{VTK}%
  \BibitemOpen
  \bibfield  {author} {\bibinfo {author} {\bibfnamefont {A.~G.}\ \bibnamefont
  {Vladimirov}}, \bibinfo {author} {\bibfnamefont {D.}~\bibnamefont {Turaev}},\
  and\ \bibinfo {author} {\bibfnamefont {G.}~\bibnamefont {Kozyreff}},\
  }\href@noop {} {\bibfield  {journal} {\bibinfo  {journal} {Opt. Lett.}\
  }\textbf {\bibinfo {volume} {29}},\ \bibinfo {pages} {1221} (\bibinfo {year}
  {2004}{\natexlab{a}})}\BibitemShut {NoStop}%
\bibitem [{\citenamefont {Vladimirov}\ and\ \citenamefont
  {Turaev}(2004)}]{VT04}%
  \BibitemOpen
  \bibfield  {author} {\bibinfo {author} {\bibfnamefont {A.~G.}\ \bibnamefont
  {Vladimirov}}\ and\ \bibinfo {author} {\bibfnamefont {D.}~\bibnamefont
  {Turaev}},\ }\href@noop {} {\bibfield  {journal} {\bibinfo  {journal}
  {Radiophys. \& Quant. Electron.}\ }\textbf {\bibinfo {volume} {47}},\
  \bibinfo {pages} {857} (\bibinfo {year} {2004})}\BibitemShut {NoStop}%
\bibitem [{\citenamefont {Vladimirov}\ and\ \citenamefont
  {Turaev}(2005)}]{VT05}%
  \BibitemOpen
  \bibfield  {author} {\bibinfo {author} {\bibfnamefont {A.~G.}\ \bibnamefont
  {Vladimirov}}\ and\ \bibinfo {author} {\bibfnamefont {D.}~\bibnamefont
  {Turaev}},\ }\href@noop {} {\bibfield  {journal} {\bibinfo  {journal} {Phys.
  Rev. A}\ }\textbf {\bibinfo {volume} {72}},\ \bibinfo {pages} {033808}
  (\bibinfo {year} {2005})}\BibitemShut {NoStop}%
\bibitem [{\citenamefont {Aadhi}\ \emph {et~al.}(2019)\citenamefont {Aadhi},
  \citenamefont {Kovalev}, \citenamefont {Kues}, \citenamefont {Roztocki},
  \citenamefont {Reimer}, \citenamefont {Zhang}, \citenamefont {Wang},
  \citenamefont {Little}, \citenamefont {Chu}, \citenamefont {Wang} \emph
  {et~al.}}]{aadhi2019highly}%
  \BibitemOpen
  \bibfield  {author} {\bibinfo {author} {\bibfnamefont {A.}~\bibnamefont
  {Aadhi}}, \bibinfo {author} {\bibfnamefont {A.~V.}\ \bibnamefont {Kovalev}},
  \bibinfo {author} {\bibfnamefont {M.}~\bibnamefont {Kues}}, \bibinfo {author}
  {\bibfnamefont {P.}~\bibnamefont {Roztocki}}, \bibinfo {author}
  {\bibfnamefont {C.}~\bibnamefont {Reimer}}, \bibinfo {author} {\bibfnamefont
  {Y.}~\bibnamefont {Zhang}}, \bibinfo {author} {\bibfnamefont
  {T.}~\bibnamefont {Wang}}, \bibinfo {author} {\bibfnamefont {B.~E.}\
  \bibnamefont {Little}}, \bibinfo {author} {\bibfnamefont {S.~T.}\
  \bibnamefont {Chu}}, \bibinfo {author} {\bibfnamefont {Z.}~\bibnamefont
  {Wang}}, \emph {et~al.},\ }\href@noop {} {\bibfield  {journal} {\bibinfo
  {journal} {Optics express}\ }\textbf {\bibinfo {volume} {27}},\ \bibinfo
  {pages} {25251} (\bibinfo {year} {2019})}\BibitemShut {NoStop}%
\bibitem [{\citenamefont {Tran}\ \emph {et~al.}(2008)\citenamefont {Tran},
  \citenamefont {Lee}, \citenamefont {Lee},\ and\ \citenamefont
  {Han}}]{tran2008switchable}%
  \BibitemOpen
  \bibfield  {author} {\bibinfo {author} {\bibfnamefont {T.~V.~A.}\
  \bibnamefont {Tran}}, \bibinfo {author} {\bibfnamefont {K.}~\bibnamefont
  {Lee}}, \bibinfo {author} {\bibfnamefont {S.~B.}\ \bibnamefont {Lee}},\ and\
  \bibinfo {author} {\bibfnamefont {Y.-G.}\ \bibnamefont {Han}},\ }\href@noop
  {} {\bibfield  {journal} {\bibinfo  {journal} {Optics express}\ }\textbf
  {\bibinfo {volume} {16}},\ \bibinfo {pages} {1460} (\bibinfo {year}
  {2008})}\BibitemShut {NoStop}%
\bibitem [{\citenamefont {Yang}\ \emph {et~al.}(2012)\citenamefont {Yang},
  \citenamefont {Zhang}, \citenamefont {Yang}, \citenamefont {Yang},
  \citenamefont {Yue},\ and\ \citenamefont {Yang}}]{yang2012chaotic}%
  \BibitemOpen
  \bibfield  {author} {\bibinfo {author} {\bibfnamefont {L.}~\bibnamefont
  {Yang}}, \bibinfo {author} {\bibfnamefont {L.}~\bibnamefont {Zhang}},
  \bibinfo {author} {\bibfnamefont {R.}~\bibnamefont {Yang}}, \bibinfo {author}
  {\bibfnamefont {L.}~\bibnamefont {Yang}}, \bibinfo {author} {\bibfnamefont
  {B.}~\bibnamefont {Yue}},\ and\ \bibinfo {author} {\bibfnamefont
  {P.}~\bibnamefont {Yang}},\ }\href@noop {} {\bibfield  {journal} {\bibinfo
  {journal} {Optics Communications}\ }\textbf {\bibinfo {volume} {285}},\
  \bibinfo {pages} {143} (\bibinfo {year} {2012})}\BibitemShut {NoStop}%
\bibitem [{\citenamefont {Yun}\ \emph {et~al.}(2012)\citenamefont {Yun},
  \citenamefont {Liu},\ and\ \citenamefont {Mao}}]{yun2012observation}%
  \BibitemOpen
  \bibfield  {author} {\bibinfo {author} {\bibfnamefont {L.}~\bibnamefont
  {Yun}}, \bibinfo {author} {\bibfnamefont {X.}~\bibnamefont {Liu}},\ and\
  \bibinfo {author} {\bibfnamefont {D.}~\bibnamefont {Mao}},\ }\href@noop {}
  {\bibfield  {journal} {\bibinfo  {journal} {Optics Express}\ }\textbf
  {\bibinfo {volume} {20}},\ \bibinfo {pages} {20992} (\bibinfo {year}
  {2012})}\BibitemShut {NoStop}%
\bibitem [{\citenamefont {Li}\ \emph {et~al.}(2014)\citenamefont {Li},
  \citenamefont {Zhang}, \citenamefont {Sun}, \citenamefont {Luo},
  \citenamefont {Liu}, \citenamefont {Yan}, \citenamefont {Mou}, \citenamefont
  {Zhang},\ and\ \citenamefont {Turitsyn}}]{li2014all}%
  \BibitemOpen
  \bibfield  {author} {\bibinfo {author} {\bibfnamefont {J.}~\bibnamefont
  {Li}}, \bibinfo {author} {\bibfnamefont {Z.}~\bibnamefont {Zhang}}, \bibinfo
  {author} {\bibfnamefont {Z.}~\bibnamefont {Sun}}, \bibinfo {author}
  {\bibfnamefont {H.}~\bibnamefont {Luo}}, \bibinfo {author} {\bibfnamefont
  {Y.}~\bibnamefont {Liu}}, \bibinfo {author} {\bibfnamefont {Z.}~\bibnamefont
  {Yan}}, \bibinfo {author} {\bibfnamefont {C.}~\bibnamefont {Mou}}, \bibinfo
  {author} {\bibfnamefont {L.}~\bibnamefont {Zhang}},\ and\ \bibinfo {author}
  {\bibfnamefont {S.~K.}\ \bibnamefont {Turitsyn}},\ }\href@noop {} {\bibfield
  {journal} {\bibinfo  {journal} {Optics express}\ }\textbf {\bibinfo {volume}
  {22}},\ \bibinfo {pages} {7875} (\bibinfo {year} {2014})}\BibitemShut
  {NoStop}%
\bibitem [{\citenamefont {Nizette}\ and\ \citenamefont
  {Vladimirov}(2021)}]{nizette2021generalized}%
  \BibitemOpen
  \bibfield  {author} {\bibinfo {author} {\bibfnamefont {M.}~\bibnamefont
  {Nizette}}\ and\ \bibinfo {author} {\bibfnamefont {A.~G.}\ \bibnamefont
  {Vladimirov}},\ }\href@noop {} {\bibfield  {journal} {\bibinfo  {journal}
  {Physical Review E, accepted for publication}\ } (\bibinfo {year}
  {2021})}\BibitemShut {NoStop}%
\bibitem [{\citenamefont {Lai}\ \emph {et~al.}(2005)\citenamefont {Lai},
  \citenamefont {Shum},\ and\ \citenamefont {Binh}}]{lai2005nolm}%
  \BibitemOpen
  \bibfield  {author} {\bibinfo {author} {\bibfnamefont {W.~J.}\ \bibnamefont
  {Lai}}, \bibinfo {author} {\bibfnamefont {P.}~\bibnamefont {Shum}},\ and\
  \bibinfo {author} {\bibfnamefont {L.}~\bibnamefont {Binh}},\ }\href@noop {}
  {\bibfield  {journal} {\bibinfo  {journal} {IEEE journal of quantum
  electronics}\ }\textbf {\bibinfo {volume} {41}},\ \bibinfo {pages} {986}
  (\bibinfo {year} {2005})}\BibitemShut {NoStop}%
\bibitem [{\citenamefont {Pimenov}\ \emph {et~al.}(2017)\citenamefont
  {Pimenov}, \citenamefont {Slepneva}, \citenamefont {Huyet},\ and\
  \citenamefont {Vladimirov}}]{pimenovprl}%
  \BibitemOpen
  \bibfield  {author} {\bibinfo {author} {\bibfnamefont {A.}~\bibnamefont
  {Pimenov}}, \bibinfo {author} {\bibfnamefont {S.}~\bibnamefont {Slepneva}},
  \bibinfo {author} {\bibfnamefont {G.}~\bibnamefont {Huyet}},\ and\ \bibinfo
  {author} {\bibfnamefont {A.~G.}\ \bibnamefont {Vladimirov}},\ }\href
  {https://doi.org/10.1103/PhysRevLett.118.193901} {\bibfield  {journal}
  {\bibinfo  {journal} {Phys. Rev. Lett.}\ }\textbf {\bibinfo {volume} {118}},\
  \bibinfo {pages} {193901} (\bibinfo {year} {2017})}\BibitemShut {NoStop}%
\bibitem [{\citenamefont {Pimenov}\ \emph {et~al.}(2020)\citenamefont
  {Pimenov}, \citenamefont {Amiranashvili},\ and\ \citenamefont
  {Vladimirov}}]{pimenov2020temporal}%
  \BibitemOpen
  \bibfield  {author} {\bibinfo {author} {\bibfnamefont {A.}~\bibnamefont
  {Pimenov}}, \bibinfo {author} {\bibfnamefont {S.}~\bibnamefont
  {Amiranashvili}},\ and\ \bibinfo {author} {\bibfnamefont {A.~G.}\
  \bibnamefont {Vladimirov}},\ }\href@noop {} {\bibfield  {journal} {\bibinfo
  {journal} {Mathematical Modelling of Natural Phenomena}\ }\textbf {\bibinfo
  {volume} {15}},\ \bibinfo {pages} {47} (\bibinfo {year} {2020})}\BibitemShut
  {NoStop}%
\bibitem [{\citenamefont {Vladimirov}\ \emph
  {et~al.}(2004{\natexlab{b}})\citenamefont {Vladimirov}, \citenamefont
  {Turaev},\ and\ \citenamefont {Kozyreff}}]{vladimirov}%
  \BibitemOpen
  \bibfield  {author} {\bibinfo {author} {\bibfnamefont {A.~G.}\ \bibnamefont
  {Vladimirov}}, \bibinfo {author} {\bibfnamefont {D.}~\bibnamefont {Turaev}},\
  and\ \bibinfo {author} {\bibfnamefont {G.}~\bibnamefont {Kozyreff}},\
  }\href@noop {} {\bibfield  {journal} {\bibinfo  {journal} {Opt. Lett.}\
  }\textbf {\bibinfo {volume} {29}},\ \bibinfo {pages} {1221} (\bibinfo {year}
  {2004}{\natexlab{b}})}\BibitemShut {NoStop}%
\bibitem [{\citenamefont {Cai}\ and\ \citenamefont {Chen}(2010)}]{cai201040}%
  \BibitemOpen
  \bibfield  {author} {\bibinfo {author} {\bibfnamefont {T.}~\bibnamefont
  {Cai}}\ and\ \bibinfo {author} {\bibfnamefont {L.~R.}\ \bibnamefont {Chen}},\
  }\href@noop {} {\bibfield  {journal} {\bibinfo  {journal} {Optics express}\
  }\textbf {\bibinfo {volume} {18}},\ \bibinfo {pages} {18113} (\bibinfo {year}
  {2010})}\BibitemShut {NoStop}%
\bibitem [{\citenamefont {Agrawal}\ and\ \citenamefont
  {Olsson}(1989)}]{agrawal1989self}%
  \BibitemOpen
  \bibfield  {author} {\bibinfo {author} {\bibfnamefont {G.~P.}\ \bibnamefont
  {Agrawal}}\ and\ \bibinfo {author} {\bibfnamefont {N.~A.}\ \bibnamefont
  {Olsson}},\ }\href@noop {} {\bibfield  {journal} {\bibinfo  {journal} {IEEE
  Journal of quantum electronics}\ }\textbf {\bibinfo {volume} {25}},\ \bibinfo
  {pages} {2297} (\bibinfo {year} {1989})}\BibitemShut {NoStop}%
\bibitem [{\citenamefont {Fermann}\ \emph {et~al.}(1990)\citenamefont
  {Fermann}, \citenamefont {Haberl}, \citenamefont {Hofer},\ and\ \citenamefont
  {Hochreiter}}]{fermann1990nonlinear}%
  \BibitemOpen
  \bibfield  {author} {\bibinfo {author} {\bibfnamefont {M.~E.}\ \bibnamefont
  {Fermann}}, \bibinfo {author} {\bibfnamefont {F.}~\bibnamefont {Haberl}},
  \bibinfo {author} {\bibfnamefont {M.}~\bibnamefont {Hofer}},\ and\ \bibinfo
  {author} {\bibfnamefont {H.}~\bibnamefont {Hochreiter}},\ }\href@noop {}
  {\bibfield  {journal} {\bibinfo  {journal} {Optics Letters}\ }\textbf
  {\bibinfo {volume} {15}},\ \bibinfo {pages} {752} (\bibinfo {year}
  {1990})}\BibitemShut {NoStop}%
\bibitem [{\citenamefont {Yanchuk}\ and\ \citenamefont
  {Wolfrum}(2010)}]{Yanchuk2010a}%
  \BibitemOpen
  \bibfield  {author} {\bibinfo {author} {\bibfnamefont {S.}~\bibnamefont
  {Yanchuk}}\ and\ \bibinfo {author} {\bibfnamefont {M.}~\bibnamefont
  {Wolfrum}},\ }\href {https://doi.org/10.1137/090751335} {\bibfield  {journal}
  {\bibinfo  {journal} {SIAM J. Appl. Dyn. Syst.}\ }\textbf {\bibinfo {volume}
  {9}},\ \bibinfo {pages} {519} (\bibinfo {year} {2010})}\BibitemShut {NoStop}%
\bibitem [{\citenamefont {Zhang}\ \emph {et~al.}(2008)\citenamefont {Zhang},
  \citenamefont {Ye}, \citenamefont {Sang},\ and\ \citenamefont
  {Nie}}]{zhang2008passively}%
  \BibitemOpen
  \bibfield  {author} {\bibinfo {author} {\bibfnamefont {Z.~X.}\ \bibnamefont
  {Zhang}}, \bibinfo {author} {\bibfnamefont {Z.~Q.}\ \bibnamefont {Ye}},
  \bibinfo {author} {\bibfnamefont {M.~H.}\ \bibnamefont {Sang}},\ and\
  \bibinfo {author} {\bibfnamefont {Y.~Y.}\ \bibnamefont {Nie}},\ }\href@noop
  {} {\bibfield  {journal} {\bibinfo  {journal} {Laser Physics Letters}\
  }\textbf {\bibinfo {volume} {5}},\ \bibinfo {pages} {364} (\bibinfo {year}
  {2008})}\BibitemShut {NoStop}%
\bibitem [{\citenamefont {Lee}\ \emph {et~al.}(2015)\citenamefont {Lee},
  \citenamefont {Koo}, \citenamefont {Jhon},\ and\ \citenamefont
  {Lee}}]{lee2015femtosecond}%
  \BibitemOpen
  \bibfield  {author} {\bibinfo {author} {\bibfnamefont {J.}~\bibnamefont
  {Lee}}, \bibinfo {author} {\bibfnamefont {J.}~\bibnamefont {Koo}}, \bibinfo
  {author} {\bibfnamefont {Y.~M.}\ \bibnamefont {Jhon}},\ and\ \bibinfo
  {author} {\bibfnamefont {J.~H.}\ \bibnamefont {Lee}},\ }\href@noop {}
  {\bibfield  {journal} {\bibinfo  {journal} {Optics express}\ }\textbf
  {\bibinfo {volume} {23}},\ \bibinfo {pages} {6359} (\bibinfo {year}
  {2015})}\BibitemShut {NoStop}%
\end{thebibliography}

\end{document}